\documentclass[proceedings]{JHEP}
\usepackage{epsfig}
\conference{TMR9: Quantum aspects of gauge theories, 
supersymmetry and unification}

\newcommand{\non}{\nonumber\\}
\def\cM{{\cal M}}
\def\cN{{\cal N}}
\def\cF{{\cal F}}
\newcommand{\be}{\begin{equation}}
\newcommand{\ee}{\end{equation}}
\newcommand{\eel}[1]{\label{#1}\end{equation}}
\newcommand{\bea}{\begin{eqnarray}}
\newcommand{\eea}{\end{eqnarray}}
\newcommand{\eeal}[1]{\label{#1}\end{eqnarray}}
\newcommand{\baq}{\begin{equation}\begin{array}{rcl}}
\newcommand{\eaq}{\end{aryray}\end{equation}}
\newcommand{\eaql}[1]{\end{array}\label{#1}\end{equation}}
\newcommand{\beac}{\begin{equation}\begin{array}{rcl}}
\newcommand{\eeacn}[1]{\end{array}\label{#1}\end{equation}}
\newcommand{\ba}{\begin{array}}
\newcommand{\ea}{\end{array}}
\newcommand{\equ}[1]{(\ref{#1})}
\newcommand{\beq}{\begin{equation}}
\newcommand{\eeq}{\end{equation}}
\newcommand{\beqar}{\begin{eqnarray}}
\newcommand{\eeqar}{\end{eqnarray}}

\title{Warped Compactifications and AdS/CFT\footnote{Talk presented at the TMR conference in Paris, September 99.}}
\author{Yaron Oz\\
Theory Division, CERN\\
CH-1211, Geneva 23, Switzerland\\}
\abstract{In this talk we discuss two classes of examples
of {\it warped products} of 
AdS spaces in the context of the AdS/CFT correspondence. 
The first class of examples appears in the construction
of  dual Type I' string descriptions to five dimensional 
supersymmetric fixed points with $E_{N_f+1}$ global symmetry. 
The background is obtained as the near horizon geometry of the D4-D8 
brane system in massive Type IIA supergravity. 
The second class of examples appears when 
considering the  $\cN=2$ superconformal
theories defined on a $3+1$ dimensional hyperplane 
intersection of two sets 
of M5 branes.
We use the dual string 
formulations to deduce properties of these field theories.}

\begin{document}

\section{Introduction}
The consideration of the near horizon geometry of branes on one hand, and
the low energy dynamics on their worldvolume on the other hand 
has lead to conjectured duality relations between field  theories and 
string theory (M theory) on certain backgrounds \cite{mal,GKP98,WittenAdS,us}.
The field theories under discussion are in various dimensions,
can be conformal or not, and with or without supersymmetry.
These properties are reflected by the type of string/M theory 
backgrounds of the dual description.

In this talk we discuss two classes of examples
of {\it warped products} of 
AdS spaces in the context of the AdS/CFT correspondence. 
The warped product structure is a fibration of 
AdS over some manifold $\cM$ \cite{Nicolai,vanNieuwenhuizen}.
It is the most general form of a metric that has the isometry of an
AdS space \cite{van}.

The first class of examples \cite{bo} appears in the construction
of  dual 
Type I' string descriptions to five dimensional supersymmetric
fixed points with $E_{N_f+1}$  global symmetry, $E_{N_f+1} =
(E_8,E_7,E_6,E_5 = \\ Spin(10), 
E_4=SU(5),
E_3=SU(3) \times SU(2), E_2=SU(2) \times U(1), 
E_1 = SU(2))$.
These fixed points are obtained in the limit of infinite bare coupling
of $\cN=2$ supersymmetric gauge theories with gauge
group $Sp(Q_4)$, $N_f < 8$ massless hypermultiplets in the 
fundamental representation and one
massless hypermultiplet in the anti-symmetric representation 
\cite{Seiberg96,IMS97}. These theories were discussed in the context of 
the AdS/SCFT correspondence in \cite{Ferrara}.
The dual background is obtained 
as the near horizon geometry of the D4-D8 brane system in massive 
Type IIA supergravity. 
The ten dimensional space 
is a fibration of $AdS_6$ over $S^4$ and has the isometry group  
$SO(2,5) \times SO(4)$.
This space provides the spontaneous compactification of
massive Type IIA supergravity in ten dimensions to the $F(4)$ 
gauged supergravity in six dimensions \cite{Romans2}.

The second class of examples \cite{ao} appears when 
considering the  $\cN=2$ superconformal
theories defined on a $3+1$ dimensional hyperplane 
intersection of two sets 
of M5 branes.
The $\cN=2$ supersymmetry algebra in four dimensions
contains a central extension term that corresponds to string charges
in the adjoint representation of the $SU(2)_R$ part of the R-symmetry group.
This implies that $\cN=2$ supersymmetric gauge theories in four dimensions
can have BPS string configurations at certain regions  in their moduli space
of vacua. 
In particular, at certain points in the moduli space of vacua these strings
can become tensionless.
A brane configuration that exhibits this phenomena consists of two  sets 
of M5 branes intersecting on $3+1$ dimensional hyperplane.
The theory on the intersection is $\cN=2$ supersymmetric.
One can stretch M2 branes between the two sets of M5 branes in a configuration
that preserves
half of the supersymmetry.
This can be viewed as a BPS string of the four dimensional theory,
and we will study such brane configurations using the AdS/CFT 
correspondence \cite{mal,us}.

The talk is organized as follows. 
In the next two sections we will discuss the first class of examples.
In section 2 we will discuss the 
D4-D8 brane system and its relation to the five dimensional fixed points.
In section 3 we will construct the dual string description and use it
to deduce some properties of the fixed points.
In the following two sections we will discuss the second class of examples.
In section 4 we will construct the 
dual string description of the four dimensional theory on the 
intersection and discuss the field theory on the intersection.
In section 5 we will use the dual string description to deduce
some properties of these strings. We will argue that from the 
four dimensional field theory viewpoint they are simply BPS string configurations
on the Higgs branch.

\section{The D4-D8 Brane System}

We start with Type I string theory on $R^9 \times S^1$ with $N$ coinciding
D5 branes wrapping the circle.
The six dimensional D5 brane worldvolume theory possesses $\cN=1$ 
supersymmetry. It has 
an $Sp(N)$ gauge group, one hypermultiplet
in the antisymmetric representation of $Sp(N)$ from the DD sector and 
16 hypermultiplets
in the fundamental representation from the DN sector. 
Performing T-duality on the circle results in Type I' theory compactified on 
the interval $S^1/Z_2$ with two orientifolds
(O8 planes) located at the fixed points. The D5 branes become
D4 branes and there are 16 D8 branes located at points on 
the interval. They cancel the -16 units of D8 brane charge carried
by the two O8 planes. The locations of the D8 branes  
correspond to masses for the hypermultiplets in the fundamental
representation arising from the open strings between the D4 branes and
the D8 branes. The  hypermultiplet
in the antisymmetric representation is massless.

Consider first $N=1$, namely one D4 brane. The worldvolume
gauge group is $Sp(1) \simeq SU(2)$. The five-dimensional
vector multiplet contains as bosonic fields the gauge field
and one real scalar. The scalar parametrizes the location of the D4 branes
in the interval, and the gauge group is broken to $U(1)$ unless the 
D4 brane is located at one of the fixed points. 
A hypermultiplet contains four real (two complex) scalars.
The $N_f$ massless matter hypermultiplets in the fundamental and the 
antisymmetric
hypermultiplet (which is a trivial representation for $Sp(1)$)
parametrize the Higgs branch of the theory. 
It is the moduli space
of $SO(2N_f)$ one-instanton. The theory has an $SU(2)_R$ R-symmetry. The 
two supercharges as well as the scalars in the hypermultiplet transform 
as a doublet under $SU(2)_R$. In addition,
the theory has a global $SU(2) \times SO(2N_f) \times U(1)_I$ symmetry.
The $SU(2)$ factor of the global symmetry group is associated with the 
massless antisymmetric hypermultiplet and is only present if $N > 1$, 
the $SO(2N_f)$ group is associated with the $N_f$ massless
hypermultiplets in the fundamental and the $U(1)_I$ part corresponds 
to the instanton number conservation.

Consider the D8 brane background metric. It takes the form
\bea\label{1} 
ds^2 &=& H_8^{-1/2} (-dt^2 + dx_1^2 + \ldots + dx_8^2) \non
&+& H_8^{1/2} dz^2 \ ,
\non
e^{-\phi} &=& H_8^{5/4} \ .
\eea
$H_8$ is a harmonic function on the interval parametrized by $z$.
Therefore $H_8$ is a piecewise linear function in $z$ where the slope
is constant between two D8 branes and decreases by one unit for each D8
brane crossed. 
Thus,
\be
H_8(z) = c + 16 \frac{z}{l_s} - \sum_{i=1}^{16} \frac{|z - z_i|}{l_s} - 
\sum_{i=1}^{16} \frac{|z + z_i|}{l_s} \ ,
\eel{d8backg}
where the $z_i$ denote the locations of the 16 D8 branes.

Denote the D4 brane worldvolume coordinates by $t,x_1 \ldots x_4$.
The D4 brane is
located at some point in $x_5\ldots x_8$ and $z$. 
We can consider it to be a probe of the D8 branes background.
The gauge coupling $g$  of the D4 brane worldvolume theory
and the harmonic function
$H_8$ are related as can be seen 
by expanding the DBI action of a D4 brane in the background \equ{d8backg}.
We  get 
\be\label{3}
g^2=\frac{l_s}{H_8} \ ,
\ee
where $g_{cl}^2 = \frac{c}{l_s}$ corresponds to the classical gauge coupling. 
In the field theory limit we take $l_s \rightarrow 0$ keeping
the gauge coupling $g$ fixed, 
thus, 
\be\label{gym}
g^2=\mbox{fixed}, ~~\Rightarrow ~~ \phi= \frac{z}{l_s^2}=
\mbox{fixed}, ~~l_s \to 0 \ .
\ee
In this limit we have
\be
\frac{1}{g^2} = \frac{1}{g_{cl}^2} + 16 \phi - \sum_{i=1}^{16}|\phi - m_i| - 
\sum_{i=1}^{16}|\phi + m_i|\ ,
\eel{coupling}
where the masses $m_i=\frac{z_i}{l_s^2}$.
Note that in the field theory limit we are studying the region
near $z=0$. The coordinate $\phi$ takes values in $R^{+}$ and 
parametrizes the 
field theory Coulomb branch.

Seiberg argued \cite{Seiberg96} that the theory at the origin
of the Coulomb branch obtained in the limit $g_{cl} = \infty$ with 
$N_f < 8$ massless hypermultiplets is a non trivial fixed point.
 The restriction $N_f < 8$ can be seen in the supergravity description
as a requirement for the harmonic function
$H_8$ in equation (\ref{d8backg}) to be positive when $c=0$
and $z_i=0$.
At the fixed point the global symmetry is enhanced
to $SU(2) \times E_{N_f+1}$.
The Higgs branch is expected to become the moduli space of  
$E_{N_f+1}$ one-instanton.

The generalization to $N=Q_4$ D4 branes is straightforward \cite{IMS97}.
The gauge group is now $Sp(Q_4)$ and the global symmetry is as before.
The Higgs branch is now the moduli space of $SO(2N_f)$ $Q_4$-instantons.
At the fixed point the global symmetry is enhanced as before and 
the Higgs branch is expected to become the moduli space of  $E_{N_f+1}$ 
$Q_4$-instanton.
Our interest in the next section will be in finding dual string 
descriptions of these fixed points.

\section{The Dual String Description}

The low energy description of our system is given by Type I' supergravity
\cite{PW}. 
The region between two D8 branes is, as discussed in  \cite{Polchinski},
described  by massive Type IIA supergravity \cite{Romans1}.  
The configurations that we will study
have $N_f$ D8 branes located at one O8 plane and $16-N_f$ D8 branes
at the other O8 plane. Therefore we are always between D8 branes and never
encounter the situation where we cross D8 branes, and the massive Type IIA 
supergravity description is sufficient.  

The bosonic part of the massive Type IIA 
action (in string frame)
including a six-form gauge field strength which is the
dual of the RR four-form field strength is
\bea\label{massiveIIA}
S &=& \frac{1}{2 \kappa_{10}^2}\int d^{10}x \sqrt{-g}
( e^{-2 \Phi}(R + 4 \partial_\mu \Phi
\partial^\mu \Phi ) \non
&-& \frac{1}{2 \cdot 6!} |F_6|^2 
- \frac{1}{2}  m^2 ) 
\eea
where the mass parameter is given by
\be
m = \sqrt{2} (8-N_f) \mu_8 \kappa_{10} = \frac{8-N_f}{2 \pi l_s} \ .
\label{m}
\ee
We use the conventions 
$\kappa_{10} = 8 \pi^{7/2} l_s^4$ and $\mu_8 = (2\pi)^{-9/2} l_s^{-5}$
for the gravitational coupling and the D8 brane charge, respectively.

The Einstein equations derived from
\equ{massiveIIA} read (in Einstein frame metric) 
\bea\label{eom}
2 R_{ij} &=& g_{ij}(R - \frac{1}{2} |\partial\Phi|^2  -
\frac{e^{-\Phi/2}}{2 \cdot 6!} |F_6|^2 - \frac{m^2}{2}
e^{5\Phi/2}) \nonumber \\ 
&+&\partial_i \Phi \partial_j \Phi + \frac{e^{-\Phi/2}}{5!}
F_{im_1 \ldots m_5} F_j^{m_1 \ldots m_5} \ , \nonumber \\
0 &=&\nabla^j \partial_j \Phi - \frac{5 m^2}{4} e^{5\Phi/2} +
\frac{e^{-\Phi/2}}{4 \cdot 6!} |F_6|^2  \ , \nonumber \\ 
0 &=& \nabla^i \left( e^{-\Phi/2} F_{i m_1 \ldots m_5} \right) \ .
\eea
Except for these formulas we will be using the string frame only.

For identifying the solutions of D4 branes localized on D8 branes 
it is more convenient to start with the conformally flat form 
of the D8 brane supergravity solution.
It takes the form  \cite{PW}
\bea
ds^2 &=& \Omega(z)^2 ( -dt^2 + \ldots + dx_4^2 \non
&+& d\tilde{r}^2 +
\tilde{r}^2 d\Omega_3^2 + dz^2 ) \ , \\
e^\Phi &=& C \left( \frac{3}{2} C m z \right)^{-\frac{5}{6}} \, 
\Omega(z) = \left( \frac{3}{2} C m z \right)^{-\frac{1}{6}} \ .
\nonumber
\eea
In these coordinates the harmonic function of $Q_4$ localized D4 branes 
in the near horizon limit derived from \equ{eom} reads
\be
H_4 = \frac{Q_4}{l_s^{10/3}(\tilde{r}^2 + z^2)^{5/3}} \ .
\label{H4}
\ee
This  localized  D4-D8 brane system  solution, in a different coordinate
system,  has been constructed in \cite{Youm1}. 
One way to determine the harmonic function (\ref{H4}) of the localized D4 
branes is to solve  the Laplace equation in the background of the D8 branes.

It is useful to make a change of coordinates $z = r \sin \alpha, 
\tilde{r}= r \cos \alpha, 0 \leq \alpha \leq \pi/2$.
We get 
\bea\label{sol}
ds^2 &=& \Omega^2 ( H_4^{-\frac{1}{2}}(-dt^2 + 
\ldots dx_4^2) \non
&+& H_4^{\frac{1}{2}} (dr^2 + r^2 d\Omega_4^2 ) ) 
\nonumber \ ,\\
e^\Phi &=& C \left( \frac{3}{2} C m r \sin \alpha \right)^{-\frac{5}{6}} 
H_4^{-\frac{1}{4}} \ , \non
\Omega &=& \left( \frac{3}{2} C m r \sin \alpha \right)^{-\frac{1}{6}} \ ,\\
F_{01234r} &=& \frac{1}{C} \partial_r \left( H_4^{-1} \right) \ ,
\nonumber
\eea
where $C$ is an arbitrary parameter of the solution \cite{PW} and
\be
d\Omega_4^2 = d\alpha^2 + (\cos \alpha)^2 d\Omega_3^2  \ .
\label{al}
\ee
The background (\ref{sol}) is a solution of the massive Type IIA supergravity 
equations (\ref{eom}).

The metric (\ref{sol}) of the D4-D8 system can be simplified to
\bea
ds^2 &=& \left( \frac{3}{2} C m \sin \alpha\right)^{-\frac{1}{3}} 
( Q_4^{-\frac{1}{2}} r^\frac{4}{3} dx_{\|}^2 \non
&+& Q_4^\frac{1}{2}
\frac{dr^2}{r^2} + Q_4^\frac{1}{2} d\Omega_4^2) \ 
\eea
where $dx_{\|}^2 \equiv  -dt^2 + \ldots + dx_4^2$.
Define now the energy coordinate $U$ by $r^2 = l_s^5 U^3$.
That this is the energy coordinate can be seen by calculating the energy
of a fundamental string stretched in the $r$ direction or by using
the DBI action as in the previous section. 
In the field theory limit, $l_s \to 0$ with the energy $U$ fixed, 
we get the metric in the form of a {\it warped product} 
\cite{vanNieuwenhuizen} of $AdS_6 \times S^4$
\bea
ds^2  &=& l_s^2 ( \frac{3}{4 \pi} C (8-N_f) \sin \alpha)^{-\frac{1}{3}} 
( Q_4^{-\frac{1}{2}} U^2 dx_{\|}^2 \non 
&+& Q_4^\frac{1}{2}
\frac{9dU^2}{4U^2} + Q_4^\frac{1}{2} d\Omega_4^2 ) \ ,
\label{warped}
\eea
and the dilaton is given by
\be
e^\Phi = Q_4^{-\frac{1}{4}} C \left(
\frac{3}{4 \pi} C (8-N_f) \sin \alpha \right)^{-\frac{5}{6}} \ .
\label{dil}
\ee

The ten dimensional space described by (\ref{warped}) is a fibration of 
$AdS_6$ over $S^4$.
It is the most general form of a metric that has the isometry of an
$AdS_6$ space \cite{van}.
The space has a boundary at $\alpha =0$ which corresponds to the location 
of the O8 plane ($z=0$).
The boundary is of the form $AdS_6 \times S^3$. 
In addition to the $SO(2,5)$ $AdS_6$ isometries, the ten dimensional 
space has also $SO(4)$ isometries associated with the spherical 
part of the metric (\ref{warped}).
In general $S^4$ has the $SO(5)$ isometry group. However, this is 
reduced due to the warped product structure. As is easily seen   
from the form of the spherical part (\ref{al}),
only transformations excluding the $\alpha$ coordinate are isometries of 
(\ref{warped}). 
We are left with an $SO(4) \sim SU(2) \times SU(2)$ isometry group. 

The two different viewpoints of the D4-D8 brane system, 
the near horizon geometry of the brane system on one hand, and
the low energy dynamics on the D4 branes worldvolume on the other hand 
suggest a duality relation. Namely,
{\it Type I' string theory compactified on the background (\ref{warped}), 
(\ref{dil}) with a 4-form flux 
of $Q_4$ units on $S^4$ is dual to an $\cN=2$ supersymmetric 
five dimensional fixed point}.
The fixed point is obtained in the limit of
infinite coupling of $Sp(Q_4)$ gauge theory 
with $N_f$ hypermultiplets in the fundamental representation
and one hypermultiplet in the antisymmetric representation, where 
$m \sim (8-N_f)$ as in (\ref{m}).
The  $SO(2,5)$ symmetry of the compactification
corresponds to the conformal symmetry of the field theory.
The $SU(2) \times SU(2)$ symmetry of the compactification
corresponds $SU(2)_R$ R-symmetry and to the $SU(2)$ global symmetry
associated with the massless hypermultiplet in the antisymmetric
representation.

At the boundary $\alpha=0$ the dilaton (\ref{dil}) blows up and Type I'
is strongly coupled. In the weakly coupled dual heterotic string description
this is seen as an enhancement of the gauge symmetry to $E_{N_f+1}$.
One can see this enhancement of the gauge symmetry in the Type I' 
description by analysing the D0 brane dynamics near the orientifold plane
\cite{Bergman,Nilles, Schwimmer}.
This means that we have  $E_{N_f+1}$ vector fields that propagate on the 
$AdS_6 \times S^3$ boundary,
as in the Horava-Witten picture \cite{HW}.
The scalar curvature of the background (\ref{warped}), (\ref{dil}) 
\be
{\mathcal{R}} l_s^2 \sim (C (8-N_f))^\frac{1}{3} Q_4^{-\frac{1}{2}} 
(\sin \alpha)^{-\frac{5}{3}}  \ ,
\label{R}
\ee
blows up at the boundary as well. 
In the dual heterotic description the dilaton is small but the curvature 
is large, too.
For large $Q_4$ there is a region, $\sin \alpha \gg  Q_4^{-\frac{3}{10}}$,
where both curvature (\ref{R}) and
dilaton (\ref{dil}) are small and thus we can trust supergravity.

The $AdS_6$ supergroup is $F(4)$. Its bosonic subgroup is 
$SO(2,5) \times SU(2)$.
Romans constructed an $\cN=4$ six dimensional gauged supergravity with 
gauge group $SU(2)$ that realizes $F(4)$ \cite{Romans2}.
It was conjectured in \cite{Ferrara} that it is related to a 
compactification of the ten dimensional massive Type IIA supergravity.
Indeed, we find that the ten dimensional background space is the warped 
product of $AdS_6$ and $S^4$ (\ref{warped}) (with $N_f=0$). 
The reduction to six dimensions can be done in two steps.
First we can integrate over the coordinate $\alpha$. 
This yields a nine dimensional space of the form $AdS_6 \times S^3$.
We can then reduce on $S^3$ to six dimensions, while gauging its isometry 
group.
Roman's construction is based on gauging an $SU(2)$ subgroup of the 
$SO(4)$ isometry group.
Generally, it has been shown in \cite{cvetic}
that the compactification of ten dimensional massive Type IIA
supergravity to six dimensional massive Type IIA supegravity
 on warped  $S^4$, is a consistent non-linear Kaluza-Klein
ansatz for the full bosonic sector of the theory.

The massive Type IIA supergravity action in the string frame
goes like 
\be
l_s^{-8} \int \sqrt{-g} e^{-2 \Phi} {\mathcal{R}} \sim Q_4^{5/2} \ ,
\ee
suggesting that the number 
of degrees of freedom goes like  $Q_4^{5/2}$ in the regime where it is 
an applicable description. Terms in the Type I' action coming from the 
D8 brane DBI action turn out to be of the same order.
Viewed from M theory point of view, we expect the corrections
to the supergravity action to go like 
$l_p^3 \sim l_s^3 e^\Phi \sim 1/Q_4$, where $l_p$ is the eleven-dimensional
Planck length. This seems to suggest that the field theory has a $1/Q_4$ 
expansion at large $Q_4$. For example, the one-loop correction
of the form 
\be
l_s^{-8} \int \sqrt{-g} l_s^6 {\mathcal{R}}^4 \sim Q_4^{1/2}
\ee
is suppressed by $Q_4^2$ compared to the tree-level action.

According to the  $AdS$/CFT correspondence the spectrum of chiral primary 
operators of the fixed point theory can be derived from the spectrum of 
Kaluza-Klein excitations of massive Type IIA supergravity on the 
background (\ref{warped}).
We will not carry out the detailed analysis here but make a few comments.
The operators fall into representations of the $F(4)$ supergroup.
As in the case of the six dimensional $(0,1)$ fixed point with $E_8$ global 
symmetry \cite{gimon} we expect $E_{N_f+1}$ neutral operators to match the 
Kaluza-Klein reduction of fields in the bulk geometry, and 
$E_{N_f+1}$ charged  operators to match the 
Kaluza-Klein reduction of fields living on the boundary.
Among the $E_{N_f+1}$ neutral operators we expect to have dimension $3k/2$
operators of the type $Tr \phi^k$ where $\phi$ is a
complex scalar in the hypermultiplet, which parametrize the Higgs branch 
of the theory.
Like in \cite{gimon} we do not expect all these operators to be in 
short multiplets.
We expect that those in long multiplets will generically receive $1/Q_4$
corrections to their anomalous dimensions.
Unlike the hypermultiplet, the vector multiplet in five dimensions is 
not a representation of the superconformal group $F(4)$. 
Therefore, we do not expect Kaluza-Klein 
excitations corresponding to neutral operators of the type 
$Tr \varphi^k$ where $\varphi$ is a real scalar in the vector multiplet,
which parametrizes the Coulomb branch of the theory. 
Among the  $E_{N_f+1}$ charged  operators we expect to have the 
dimension four $E_{N_f+1}$ global symmetry currents that couple to the 
massless $E_{N_f+1}$ gauge fields on the boundary.

\section{The M5-M5' Brane System}

We denote the eleven dimensional space-time coordinates by $(x_{||},\vec{x},\vec{y},\vec{z})$,
where $x_{||}$ parametrize the $(0,1,2,3)$ coordinates,
$\vec{x}=(x_1,x_2)$ the $(4,5)$ coordinates, 
$\vec{y} = (y_1,y_2)$ the $(6,7)$ coordinates and 
$\vec{z}=(z_1,z_2,z_3)$ the $(8,9,10)$ coordinates.

Consider two sets of fivebranes in M theory: $N_1$ coinciding M5 branes and $N_2$ 
coinciding M5' branes.
Their worldvolume coordinates are  $M5: (x_{||},\vec{y})$
and $M5': (x_{||},\vec{x})$. 
Such a configuraion preserves eight supercharges.
The eleven-dimensional supergravity background takes the form \cite{PT,T,GKT}
\bea
ds^2_{11} &=& (H_1H_2)^{2/3}[(H_1H_2)^{-1}dx^2_{||}
\non 
&+& H^{-1}_2 d\vec{x}^2+H^{-1}_1 d\vec{y}^2+ d\vec{z}^2] \ ,
\label{m5m5}
\eea
with the 4-form field strength $\cF$
\bea
\cF = 3(*dH_1\wedge dy^1 \wedge dy^2 + *dH_2\wedge dx^1 \wedge dx^2) \ ,
\eea
where $*$ defines the dual form in the three dimensional space $(z_1,z_2,z_3)$.
When the M5 and M5' branes are only localized along the overall transverse directions $\vec{z}$
the  harmonic functions are $H_1 = 1 + l_p N_1/2|\vec{z}|,H_2 = 1 + l_p N_2/2|\vec{z}|$,
where $l_p$ is the eleven dimensional Planck length. This case has been discussed in
\cite{HK}.
The near horizon geometry in this case does not have the AdS isometry group and thus cannot
describe a dual SCFT. Moreover, there does not seem to be in this case
a theory on the intersection which decouples from the bulk physics.

Consider the semi-localized case when the M5 branes are completely localized while the
M5' branes are only localized along the overall transverse directions.  
When the branes are at the origin of $(\vec{x},\vec{z})$ 
space the harmonic functions in the near core limit of the M5' branes take the form
\cite{Youm2,L} 
\beq
H_1=1+\frac{4\pi l_p^4 N_1 N_2}{(|\vec{x}|^2+
2l_p N_2|\vec{z}|)^2}, \ \ \  
H_2=\frac{l_p N_2}{2|\vec{z}|} \ .
\label{m5m5harm}
\eeq
The numerical factors in (\ref{m5m5harm})
are determined by the requirment that the integral of ${\cal F}$ yields
the appropriate charges.
This solution can also be obtained from the localized D2-D6 brane solution of
\cite{ity} by a chain of dualities. 

It is useful to make a change of coordinates
$l_p z= (r^2 \sin\alpha^2)/2N_2, x=r \cos\alpha, 0 \leq \alpha \leq \pi/2$.
In the
near-horizon limit we want to keep the energy $U=\frac{r}{l_p^2}$ fixed.
This implies, in particular, that membranes stretched between the M5 snd M5' branes
in the $\vec{z}$ direction have finite tension $|\vec{z}|/l_p^3$.
Since we are smearing over the $\vec{y}$ directions we should keep $\vec{y}/l_p$ 
fixed. It is useful to make 
a change of variables  $\vec{y}/l_p \rightarrow \vec{y}$.
The near horizon metric is of the form of a $warped\,\,\, product$ of $AdS_5$ 
and a six dimensional manifold $\cM_6$
\bea
ds^2_{11} &=& l_p^2 (4 \pi N_1)^{-1/3}(\sin^{2/3}\alpha)(
\frac{U^2}{N_2} 
dx^2_{||} \non &+& \frac{4\pi N_1}{U^2} dU^2+ d\cM_6^2) \ , 
\label{metric11}
\eea
with
\bea
d\cM_6^2 &=& 
4\pi N_1 (d\alpha^2+\cos^2\alpha d\theta^2 
+\frac{\sin^2\alpha}{4}d\Omega_2^2 ) \non
&+&\frac{N_2}{\sin^2 \alpha}(dy^2+y^2d\psi^2) \ .
\label{M}
\eea
The metric (\ref{metric11}) has the $AdS_5$ isometry group \cite{van}. Therefore, in the
spirit
of \cite{mal},  M theory on
the background  (\ref{metric11}), (\ref{M})
should be dual to a four dimensional $\cN=2$ SCFT.

Note that the curvature of the metric diverges for small $\alpha$ as
\beq
{\cal R}\sim \frac{1}{l_p^2N_1^{2/3}
\sin^{8/3}\alpha} \ .
\eeq
Away from $\alpha=0$, eleven-dimensional supergravity can be trusted for large $N_1$.  
The singularity at $\alpha=0$ is interpreted as a signal that some degrees of freedom
have been effectively integrated and are needed in order to resolve the singularity.
These presumably correspond to membranes that end on the M5' branes.


The above M5-M5' brane system can be understood as up lifting to eleven dimensions of an elliptic brane
system of Type IIA. It consists of
$N_2$ NS5-branes with worldvolume coordinates $(0,1,2,3,4,5)$ 
periodically arranged in the 6-direction and 
$N_1$ D4-branes with worldvolume coordinates $(0,1,2,3,6)$ 
stretched between them as in figure 1. 
When we lift this brane configuration to eleven dimensions, we delocalize in the eleven 
coordinate (which in our notation is 7) and since we have delocalization in coordinate
6 as well,
we end up with the semi-localized  M5-M5' brane system.

\begin{figure}[htb]
\begin{center}
\epsfxsize=2.5in\leavevmode\epsfbox{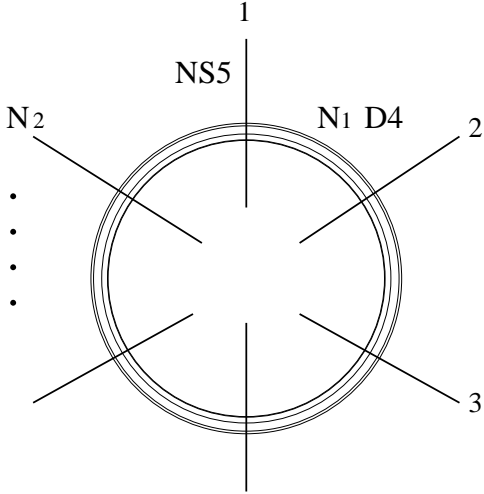}
\end{center}
\caption{ The Type IIA elliptic brane
system corresponding to the semi-localized M5-M5' brane configuration. It consists of
$N_2$ NS5-branes with worldvolume coordinates $(0,1,2,3,4,5)$ 
periodically arranged in the 6-direction and 
$N_1$ D4-branes with worldvolume coordinates $(0,1,2,3,6)$ 
stretched between them. 
}
\label{ell}
\end{figure}

The four dimensional theory at low energies on the
D4-branes worldvolume 
is an $SU(N_1)^{N_2}$ gauge theory with matter in the bi-fundamentals.
The metric (\ref{metric11}) can be viewed
as providing the 
eleven-dimensional supergravity description
of an M5 brane with worldvolume $R^4\times \Sigma$  \cite{witten}, where $\Sigma$  
is the 
Seiberg-Witten holomorphic curve (Riemann surface) associated with
the four dimensional SCFT at the origin of
the moduli space of vacua \cite{FS}.
In this brane set-up the R-symmetry group $SU(2)_R \times U(1)_R$ is realized as 
the rotation group $SU(2)_{8910} \times U(1)_{45}$. The dimensionless gauge coupling
of each $SU(N_1)$ part of the gauge group is $g_{YM}^2 \sim g_s N_2$.

The ten dimensional metric describing the elliptic Type IIA brane configuration
is  
\bea
ds^2_{10} &=& H^{-1/2}_4 dx_{||}^2+
H^{1/2}_4 d\vec{x}^2 
+H^{-1/2}_4H_{NS}dy^2 \non
&+& H^{1/2}_4H_{NS}d\vec{z}^2 \ ,
\label{dpns}
\eea
where 
\begin{equation}
H_4=1+{{4\pi g_sl_s^4 N_1N_2}\over{[|\vec{x}|^2+
2 l_s N_2|\vec{z}|]^{2}}},\ \ \ 
H_{NS}={{l_sN_2}\over {2|\vec{z}|}} \ .
\label{dpnsharm}
\end{equation}
Its near-horizon limit is
\be
ds^2_{10}= l_s^2 \left (\frac{U^2}{R^2}dx^2_{||}+
R^2\frac{dU^2}{U^2}+ d\cM_5^2 \right ) \ ,
\ee
where
\bea 
d\cM_5^2 &=& 
R^2(d\alpha^2+\cos^2\alpha d\theta^2
+{\sin^2\alpha\over 4}d\Omega_2^2) \non
&+& {N_2^2\over {R^2 \sin^2\alpha}}dy^2 \ ,
\label{metric10}
\eea
and $R^2=(4\pi g_sN_1N_2)^{1/2}$.
The curvature of the metric is 
${\cal R}\sim  \frac{1}{l_s^2R^2}$ and it has a dilaton 
$e^{\Phi}={{g_sN_2}\over {R}}\sin\alpha^{-1}$.
This supergravity solution can be trusted when both the curvature and the 
dilaton are small. Away from $\alpha=0$ we need large $N_2$ and 
$N_2 \ll N_1^{1/3}$. 
When the latter condition is not satisfied the dilaton is large and  we should consider
the eleven-dimensional description.

The supergravity solution with both the M5 branes and the M5' branes being completely localized
has not been constructed yet.
This fully localized system has been studied via a matrix description in \cite{KOY}.
In the following we make a few comments on this branes configuration.
It can be viewed as a single M5 brane 
with worldvolume $R^4\times \Sigma$, where $\Sigma$  
is a singular Riemann surface defined
by the complex equation 
\beq
x^{N_2}y^{N_1} =0 \ ,
\label{ceq}
\eeq
 where $x=x_1+ix_2, y =y_1+iy_2$.
Consider first the case $N_1=N_2=1$. The complex equation
$xy =0$
can be written by a change of variables as  
$w_1^{2} + w_2^{2} =0$. This equation describes the
singular Riemann surface at the point in the Seiberg-Witten solution
\cite{SWitten} where a monopole becomes massless.
The massless spectrum consists of a $U(1)$ vector field and one
hypermultiplet and the theory at this point  is free in the IR.
Indeed, the chiral ring associated with the singular variety $xy =0$
consists only of the identity operator.
Since the field theory is free in the IR we expect the dual string description
to be strongly coupled. That means that the dual background is highly curved and 
the classical supergravity analysis cannot be trusted.
This theory has a dual string formulation \cite{GKPelc}.
The more general $N_1,N_2$ case is harder to analyse. The reason being that
the singular locus of the variety (\ref{ceq}) is not isolated 
and the notion of a chiral ring which is used for
an isolated singularity is not appropriate.
We do expect a non trivial conformal field theory in this case, but it has not been
identified yet.

\section{BPS String Solitons}

The centrally extended $\cN=2$ supersymmetry algebra in four dimensions
takes the form \cite{FP}
\beqar
\{Q_{\alpha}^A, \bar{Q}_{\dot{\alpha}B} \} &=& \sigma_{\alpha\dot{\alpha}}^{\mu}P_{\mu}\delta_B^A
+ \sigma_{\alpha\dot{\alpha}}^{\mu}Z_{\mu B}^A,
\ , \nonumber\\
\{Q_{\alpha}^{A}, Q_{\beta}^B \} &=& \varepsilon_{\alpha\beta}Z^{[AB]} +
 \sigma_{\alpha\beta}^{\mu\nu}Z_{\mu\nu}^{(AB)} \ ,
\label{susy}
\eeqar
where $A,B=1,2$ and $Z_{\mu A}^A = 0$.
The $SU(2)_R$ part of the R-symmetry group  acts on the indices $A,B$.
In addition to the particle charge $Z^{[AB]}$, there are in (\ref{susy})
the string charges $Z_{\mu B}^A$ in the adjoint representation of $SU(2)_R$  
and the membrane charges $Z_{\mu\nu}^{(AB)}$ in the 2-fold symmetric representation
of  $SU(2)_R$.
Thus,  $\cN=2$ supersymmetric gauge theories in four dimensions
can have BPS strings and BPS domain walls in addition
to the well studied BPS particles.

The BPS particles and BPS strings are realized by stretching M2 brane between the
M5 and M5' branes.
When the membrane worldvolume coordinates are $(0,x,y)$ where $x$ is one of the $\vec{x}$
components and $y$ is one of the $\vec{y}$ components we get a BPS particle
of the four dimensional theory.
It is charged under $U(1)_R$ that acts on $\vec{x}$ and is a singlet
under $SU(2)_R$ that acts on $\vec{z}$. 
The BPS particles
exist on the Coulomb branch of the gauge theory on which $U(1)_R$ acts, as in figure 3, 
and their mass is given by the Seiberg-Witten solution.

When the membrane worldvolume coordinates are $(0,1,z)$ where $z$ is one of the $\vec{z}$
components  we get a BPS string of the four dimensional theory.
It is not charged under $U(1)_R$ and it transforms in the adjoint 
$SU(2)_R$ since $\vec{z}$ transforms in this representation. The BPS strings
exist on the Higgs branch of the gauge theory on which $SU(2)_R$ acts, as in figure 3.
The tension of the BPS string is given by $\frac{|\vec{z}|}{l_p^3}$ where $|\vec{z}|$
is the distance between the M5 and M5' branes and it is finite in the field theory
limit.
It is easy to see that we do not have in this set-up BPS domain walls since stretching a
membrane with worldvolume coordinates $(0,1,2)$ breaks the supersymmetry completely.
One expects BPS domain wall configurations when the moduli space
of vacua has disconnected components.

In the Type IIA picture the BPS string is constructed by stretching a D2 brane between
the D4 branes and NS5-brane.
The seperation between these two types of branes in the $\vec{z}$ direction
has two interpretations  
depending on whether the four dimensional gauge group has
a $U(1)$ part or not \cite{GK}. If is does then the separation is interpreted as 
the $SU(2)_R$ triplet FI parameters
$\vec{\zeta}$. The  BPS string tension is proportional to $|\vec{\zeta}|$.
If the gauge group does not have a $U(1)$ part the  
seperation between these two types of branes in the $\vec{z}$ direction is 
interpreted as giving a vev to an $SU(2)_R$ triplet component
of the meson $\tilde{Q}Q$ which transforms under $SU(2)_R$ as 
${\bf 2} \times {\bf 2} = {\bf 3}\oplus{\bf 1}$.
The BPS string tension is proportional to this vacuum expectation value.
At the origin of the moduli space, where we have SCFT, the string becomes tensionless.

\begin{figure}[htb]
\begin{center}
\epsfxsize=3in\leavevmode\epsfbox{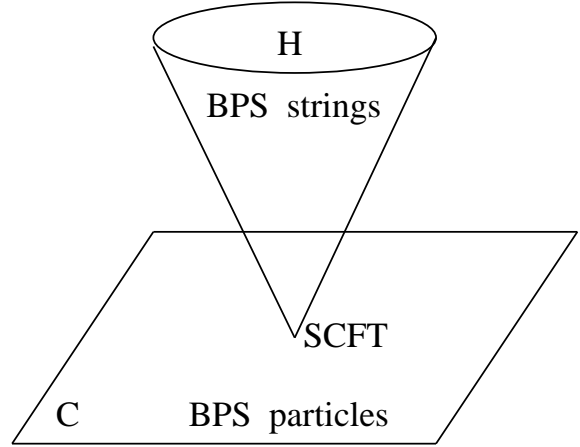}
\end{center}
\caption{There are BPS particles on the Coulomb branch and BPS
strings on the Higgs branch. The SCFT is at the intersection of these two branches. 
}
\label{ch}
\end{figure}

We can use the AdS/CFT correspondence in order to compute the self energy of 
a string and a potential between two such strings of opposite
orientation \cite{maldacena}. 
This is obtained, in the supergravity approximation, by a minimization
of the M2 brane action 
\be
S={1\over {(2\pi)^2 l^3_p}} \int d\tau d\omega d\sigma
\sqrt{det G_{\mu\nu}\partial_a x^{\mu}\partial_b X^{\nu}}
\ee 
in the background (\ref{metric11}), (\ref{M}).
Consider a static configuration $x^0=\tau,\,\, x^1=\omega$ and $x^i=x^i(\sigma)$.
Then, the energy per unit length of the string is given by
\be
E=\int{U\sin\alpha \over {4\pi^2 N_2}} \sqrt{dU^2 +U^2 d\alpha^2} \ .
\ee
Thus, the string self-energy 
\be 
E= {\cal E}(\alpha){U^2\over N_2} \sim {\cal E}(\alpha) \frac{N_1}{(\delta x_{||})^2} \ .
\label{se}
\ee
The function ${\cal E}$ parametrizes the dependence of the self-energy on the coordinate
$\alpha$. From the field theory point of view, this parametrizes 
a dependence on the moduli that parametrize the space of vacua. $\delta x_{||}$ is a cut-off.
In (\ref{se}) we used the
holographic relation between a distance  
$\delta x_{||}$ in field theory and the coordinate $U$ which in our case reads
\be
\delta x_{||} \sim \frac{(N_2N_1)^{1/2}}{U}  \ .
\label{hol} 
\ee   
This corresponds to the familiar relation \cite{SW}
\be
\delta x_{||} \sim \frac{(g_{YM}^2N_1)^{1/2}}{U} \ .
\ee

Similarly, the potential energy per units length
between two strings of opposite orientation seperated by distance $L$ is
\be
E \sim -{N_1\over L^2} \ .
\label{pot}
\ee
here $L$ is distance between two strings. 
The results (\ref{se}),(\ref{pot}) do not depend on $N_2$ which means that they
do not depend on the gauge coupling of the four dimensional 
field theory $g_{YM}^2 \sim N_2$. This is not unexpected.
The gauge coupling can be viewed as a vacuum expectation value  of a scalar
in the vector multiplet and does not appear in the hypermultiplet metric.
Since the BPS strings exist
on the Higgs branch it is natural that their potential and self-energy do not depend
on 
the gauge coupling. However, (\ref{se}) and (\ref{pot}) are only the large $N_1$ results
and will presumably have $1/N_1$ corrections.

As noted above, the theory on the intersection
of two sets of M5 branes containes BPS particles, which arise from
M2 branes stretched in the directions $(0,x,y)$.
Minimization of the M2 brane action in this case yields the potential between
these two such objects of opposite charge separated by distance $L$  
\be
V\sim {(N_1N_2)^{1/2}\over L} \ .
\ee
Again, this is expected since for the four dimensional field theory it reads as
$V\sim \frac{(g_{YM}^2N_1)^{1/2}}{L}$.

The 11-dimentional supergravity action goes like
\be
l_p^{-9}\int \sqrt{-g} {\cal R}\sim N_1^2 N_2
\ee
suggesting that the number of degrees of freedom goes like $N_1^2 N_2$. 
This is also deduced from the two-point function of the stress energy tensor.
This is, of course,  expected 
for $SU(N_1)^{N_2}$ gauge theory. 
It is curious to note that there is
similar growth ($N^3$) of the entropy for a system of $N$ {\it parallel}
M5 branes \cite{KT}.

The one loop
correction has the following form
$l_p^{-9}\int \sqrt{-g}l_p^6 {\cal R}^4\sim  N_2$
which is suppressed by $N_1^2$ compared to the tree level action
suggesting that the field theory has a $1/N_1$ expansion, which again is in agreement
with the field theory expectation.

The Type IIA background (\ref{metric10}) is T-dual to Type IIB on $AdS_5 \times S^5/Z_{N_2}$
\cite{FS}. The latter is the dual description
of the $Z_{N_2}$ orbifold of $\cN=4$ theory \cite{KS}. 
For instance the number of degrees of freedom can be understood as $c(\cN=4)/N_2 \sim N_1^2N_2$.
We can identify the spectrum of chiral primary operators with the supergravity
Kaluza-Klein excitations as analysed in \cite{OT,Gukov}.

\vskip .2in
\noindent
\acknowledgments
We would like to thank M. Alishahiha and A. Brandhuber for collaboration
on the work presented in this talk.


\newpage

\begingroup\raggedright\endgroup

\end{document}